\documentclass[aps,showpacs,preprintnumbers,amsmath,amssymb,twocolumn]{revtex4}

\usepackage{multirow}
\usepackage{verbatim}

\input{epsf}

  \def\mn{{\mu\nu}} 

\def\beq{\begin{equation}} \def\eeq{\end{equation}}
\def\beqa{\begin{eqnarray}} \def\eeqa{\end{eqnarray}}

\def\mn{{\mu\nu}}

\begin{document}

\title{Energy Conditions and Stability in $f(R)$ theories of gravity with non-minimal coupling to matter}

\author{Orfeu Bertolami \footnote{Also at Instituto de Plasmas e Fus\~ao Nuclear, Instituto Superior T\'ecnico, Lisboa\\}}
\email{orfeu@cosmos.ist.utl.pt}
\affiliation{Instituto Superior T\'ecnico, Departamento de F\'\i sica\\
             Av. Rovisco Pais 1, 1049-001 Lisboa, Portugal}

\author{Miguel Carvalho Sequeira}
\email{miguel.sequeira@ist.utl.pt}
\affiliation{Instituto Superior T\'ecnico, Departamento de F\'\i sica\\
             Av. Rovisco Pais 1, 1049-001 Lisboa, Portugal}

\date{\today}

\begin{abstract}

Recently, in the context of $f(R)$ modified theories of gravity, a new type of model has been proposed where one directly couples the scalar curvature to matter.  As any model in $f(R)$ theory, there are certain conditions which have to be satisfied in order to ensure that the model is viable and physically meaningful. In this paper, one considers this new class of models with curvature-matter coupling and study them from the point of view of the energy conditions and of their stability under the Dolgov-Kawasaki criterion.
\end{abstract}

\preprint{DF/IST-2-2009}

\maketitle


\section{Introduction}

Recently, there has been a revival of interest in gravity models where the linear scalar curvature term of the Einstein-Hilbert action is replaced by a function, $f(R)$, of the scalar curvature. This interest is due to the possibility of accounting for the late time accelerated expansion of the universe without the need of explicit additional degrees of freedom such as dark energy and also to possibly replace dark matter at galactic level and beyond (see, for instance, Refs. \cite{Nojiri06,Sotiriou:2008rp} for reviews).

In this paper, one examines models where, besides the extension in the geometric part of the action, there is also a direct coupling of a new function of the Ricci scalar to the matter Lagrangian density  \cite{Bertolami:2007} (see also Ref. \cite{Bertolami:2008zh} for a recent review):
\begin{equation} \label{action}
S=\int \left[ \frac{k}{2}f_1(R)+f_2(R){\cal L}_m \right] \sqrt{-g}d^4x\,,
\end{equation}
where $k$ is a coupling constant, $f_i(R)$ (with $i=1,2$) are arbitrary functions of the
curvature scalar $R$ and ${\cal L}_{m}$ is the Lagrangian density
corresponding to matter. One sets $k=1$ hereafter. The second function, $f_2(R)$, is usually considered to have the following form,

\begin{equation} \label{f2}
f_2(R)=1+\lambda\varphi_2(R)
\end{equation}
where $\lambda$ is a constant and $\varphi_2$ is another function of $R$.  Notice that this non-minimal curvature-matter coupling has been previously considered in order to address the issue of the accelerated expansion of the universe \cite{Odintsov}.

Varying action (\ref{action}) with respect to the metric $g_{\mu \nu }$
yields the field equations
\begin{equation}\label{fieldEq}
(f'_1+2 {\cal L}_m f'_2)R_{\mu \nu }-\frac{1}{2}f_1g_{\mu \nu }-\Delta_{\mu\nu}(f'_1+2{\cal L}_m f'_2)= f_2T_{\mu\nu }
 \,,
\end{equation}
where the prime denotes the differentiation with respect to $R$, $\Delta_{\mu\nu}=\nabla_\mu \nabla_\nu - g_{\mu\nu}\Box$ and $\Box = g^{\mu\nu}\nabla_\mu \nabla_\nu$. The matter energy-momentum tensor is defined as usual
\begin{equation}
T_{\mu \nu}=-\frac{2}{\sqrt{-g}}{\delta(\sqrt{-g}{\cal L}_m)\over \delta(g^{\mu\nu})} \,. \label{defSET}
\end{equation}

This theory exhibits interesting features. The most intriguing is the fact that the motion of a point-like test particle is non-geodesic due to the appearance of an extra force in the theory \cite{Bertolami:2007}. Actually, this issue has been a lively point of discussion in the literature \cite{Sotiriou08,Sotiriou:2008it,Bertolami:2008ab,Puetzfeld:2008xu,Bertolami:2008im}. These modified gravity models have also been examined from the point of view of their impact on stellar stability \cite{Bertolami:2007vu}.

In this work one analyzes these new modified models of gravity with non-minimal coupling between curvature and matter from the point of view of the energy conditions and their stability under the Dolgov-Kawasaki criterion.

As one shall see, the energy conditions allow one to establish under which conditions gravity remains attractive, along with the demands that the energy density is positive and cannot flow faster that light, in the context of the new modified models, while the Dolgov-Kawasaki stability criterion helps one ruling out some classes of modified models \cite{Faraoni}.

This paper is organized as follows. In section \ref{energyConditions} one investigates the energy conditions suitable for the modified gravity models with non-minimal curvature-matter coupling. One deduces an equivalent energy-momentum tensor, along with an energy density, pressure and the gravitational coupling and thus obtain the energy conditions for the modified gravity models. At the end of section \ref{energyConditions}, one illustrates the energy conditions by applying them to a fairly broad class of models. In section \ref{massInstability} one studies the generalized Dolgov-Kawasaki criterion for some modified gravity models and examine their viability in what respects their stability. Section \ref{conclusions} contains our conclusions.


\section{Energy Conditions}\label{energyConditions}

\subsection{The Raychaudhuri Equation}\label{raychaudhuriEq}

The origin of the null energy condition (NEC) and of the strong energy conditions (SEC) is the Raychaudhuri equation together with the requirement that the gravity is attractive for a space-time manifold endowed with a metric $g_\mn$ (see e.g. Ref. \cite{Hawking:1973uf} for a discussion). From the Raychaudhuri equation one has that the temporal variation of the expansion of a congruence defined by the vector field $u^\mu$ is given for the case of a congruence of timelike geodesics by
\begin{equation}\label{rayTimelike}
\frac{d \theta}{d\tau}=-\frac{1}{3}\theta^2-\sigma_{\mu\nu}\sigma^{\mu\nu}+\omega_{\mu\nu}\omega^{\mu\nu}-R_{\mu\nu}u^\mu u^\nu ~~.
\end{equation}
In the case of a congruence of null geodesics defined by the vector field $k^\mu$, the Raychaudhuri equation is given by
\begin{equation}\label{rayNulllike}
\frac{d \theta}{d\tau}=-\frac{1}{2}\theta^2-\sigma_{\mu\nu}\sigma^{\mu\nu}+\omega_{\mu\nu}\omega^{\mu\nu}-R_{\mu\nu}k^\mu k^\nu \, ,
\end{equation}
where $R_{\mu\nu}$, $\theta$, $\sigma_{\mu\nu}$ and $\omega_{\mu\nu}$ are the Ricci tensor, the expansion parameter, the shear and the rotation associated to the congruence, respectively.

It is important to realize that the Raychaudhuri equation is purely geometric and independent of the gravity theory under consideration. The connection with the gravity theory comes from the fact that, in order to relate the expansion variation with the energy-momentum tensor, one needs the field equations to obtain the Ricci tensor. Thus, through the combination of the field equations and the Raychaudhuri equation, one can set physical conditions for the energy-momentum tensor. The requirement that gravity is attractive imposes constraints on the energy-momentum tensors and 
establishes which ones are compatible. Of course, this requirement may not hold at all instances. Indeed, a repulsive interaction is what
is needed to avoid singularities as well as to achieve inflationary conditions, and to account the observed accelerated expansion of the universe.

Since $\sigma_{\mu\nu}\sigma^{\mu\nu}\geq0$, one has, from Eqs. (\ref{rayTimelike}) and (\ref{rayNulllike}), that the conditions for gravity to remain attractive ($d\theta/d\tau<0$) are given by
\begin{eqnarray}
R_{\mu\nu}u^\mu u^\nu\geq 0 \text{\qquad \qquad SEC} \label{SEC} \\
R_{\mu\nu}k^\mu k^\nu \geq 0 \text{\qquad \qquad NEC}\label{NEC}
\end{eqnarray}
for any hypersurface of orthogonal congruences ($\omega_{\mu\nu}=0$).

For the sake of completeness, one can easily see that if one inserts Einstein's field equations into Eq. (\ref{SEC}) one gets,
\begin{equation}\label{SEC2}
\left( T_{\mu\nu}-\frac{T}{2}g_{\mu\nu}\right)u^\mu u^\nu \geq 0
\end{equation}
where $T^{\mu\nu}$ is the energy-momentum tensor, $T$ is its trace and one has used units where $8\pi G=c=1$ (which shall be adopted hereafter). If one considers a perfect fluid with energy density $\rho$ and pressure $p$,
\begin{equation}\label{perfectEMTensor}
T_\mn=(\rho + p)u_\mu u_\nu - p g_\mn \, ,
\end{equation}
then Eq. (\ref{SEC2}) turns into the well known SEC for Einstein's theory,
\begin{equation}\label{SECGR}
\rho+3p\geq 0 \,.
\end{equation}

In the same fashion, if one inserts the Einstein's field equations into Eq. (\ref{NEC}), one obtains
\begin{equation}
T_\mn k^\mu k^\nu \geq 0 \,,
\end{equation}
which, after assuming Eq. (\ref{perfectEMTensor}), turns into the familiar NEC of general relativity
\begin{equation}\label{NECGR}
\rho+p \geq 0 \,.
\end{equation}
%
\subsection{Effective Energy-Momentum Tensor}

In what follows, one generalizes to the gravity model with non-minimal curvature-matter coupling the procedure developed in Ref. \cite{Santos} for $f(R)$ theories.

Rewriting Eq. (\ref{fieldEq}) in order to make explicit the Einstein tensor, $G_\mn$, one gets
\begin{equation}\label{fieldEq2}
G_\mn = \frac{f_2}{f'_1+2f'_2 {\cal L}_m}\left( \hat T_\mn+T_\mn \right) \,,
\end{equation}
where the effective energy-momentum tensor $\hat T_\mn$ has been defined as
\begin{equation}\label{effectiveEMTensor}
\hat T_\mn=\frac{1}{2}\left(\frac{f_1}{f_2} - \frac{f'_1+2{\cal L}_m f'_2}{f_2}R \right)g_\mn+\frac{1}{f_2}\Delta_\mn\left(f'_1+2{\cal L}_m f'_2\right).
\end{equation}

>From Eq. (\ref{fieldEq2}) one can define an effective coupling
\begin{equation}
\hat k=\frac{f_2}{f'_1+2{\cal L}_m f'_2} \, .
\end{equation}
The field equations can be written as
\begin{equation}
G_\mn=\hat k \left(\hat T_\mn + T_\mn \right)\,.
\end{equation}

Thus, in order to keep gravity attractive, besides conditions Eqs. (\ref{SEC})-(\ref{NEC}), $\hat k$ has to be positive from which follows the additional condition
\begin{equation}\label{couplingIneq}
\frac{f_2}{f'_1+2{\cal L}_m f'_2} >0 \, .
\end{equation}
It is important to realize that this condition is independent from the ones derived from the Raychaudhuri equation (Eqs. (\ref{SEC}) and (\ref{NEC})). While the latter is derived directly from geometric principles, the former is related with the fairly natural definition of an effective gravitational coupling.

The effective energy-momentum tensor defined by Eq. (\ref{effectiveEMTensor}) can be written in the form of a perfect fluid (Eq. (\ref{perfectEMTensor})) if one defines an effective energy density and an effective pressure. However, given the presence of the higher order derivatives in Eq. (\ref{effectiveEMTensor}), in order to proceed it is necessary to specify the metric of the space-time manifold of interest. Since one is interested in cosmological applications, the Robertson-Walker (RW) metric is a natural choice. In what follows, one considers the homogeneous and isotropic flat RW metric with the signature $(+,-,-,-)$,
\begin{equation}\label{RWmetric}
ds^2=dt^2-a^2(t) ds^2_3 \, ,
\end{equation}
where $ds^2_3$ contains the spacial part of the metric and $a(t)$ is the scale factor.

Using this metric the higher order derivative term is given by
\begin{align}
\Delta_\mn h(R,{\cal L}_m) &=(\nabla_\mu\nabla_\nu -g_\mn\Box) h(R,{\cal L}_m) \nonumber \\
					 &=(\partial_\mu\partial_\nu - g_\mn \partial_0\partial_0)h - (\Gamma^0_\mn + g_\mn 3H)\partial_0h
\end{align}
where $h(R,{\cal L}_m)$ is a generic function of $R$ and ${\cal L}_m$, $H=\dot a/a$ is the Hubble expansion parameter
and $\Gamma^0_\mn=a\dot a \delta_\mn$ (with $\mu, \nu\neq0$) are the components of the affine connection. For the flat RW metric one has $R=-6(H^2+\frac{\ddot a}{a})$. Thus, the effective energy density is given by
\begin{equation}
\hat \rho=\frac{1}{2}\left(\frac{f_1}{f_2}-\frac{f'_1+2{\cal L}_m f'_2}{f_2}R\right)-3H\frac{f''_1+2{\cal L}_mf''_2}{f_2}\dot R
\end{equation}
while the effective pressure is given by
\begin{eqnarray}
\hat p=-\frac{1}{2}\left(\frac{f_1}{f_2}-\frac{f'_1+2{\cal L}_m f'_2}{f_2}R\right)+ \nonumber \\
 (\ddot R+2H\dot R)\frac{f''_1+2{\cal L}_mf''_2}{f_2}+
\frac{f'''_1+2{\cal L}_m f'''_2}{f_2}\dot R^2 \,,
\end{eqnarray}
where the dot refers to derivative with respect to time.

\subsection{Strong and Null Energy Conditions}

>From the above results, it is straightforward to obtain the energy conditions. From Eq. (\ref{fieldEq2}) the Ricci tensor can be written as
\begin{equation}\label{Ricci}
R_\mn=\hat k \left(\hat T_\mn +T_\mn -\frac{1}{2}g_\mn(\hat T+T)\right) \ .
\end{equation}
Inserting this into Eq. (\ref{SEC}) one obtains for SEC
\begin{equation}\label{SEC3}
\hat k \left(\hat T_\mn +T_\mn\right)u^\mu u^\nu -\frac{1}{2}\hat k\left(\hat T+T \right) \geq 0 \,.
\end{equation}

Using Eq. (\ref{couplingIneq}) and assuming a perfect fluid as a matter source, one gets

\begin{eqnarray}\label{modifiedSEC}
\rho +3p - \left(\frac{f_1}{f_2}-\frac{f'_1+2{\cal L}_m f'_2}{f_2}R \right) \nonumber \\
 +3(\ddot R + H\dot R) \frac{f''_1+2{\cal L}_m f''_2}{f_2}+\nonumber \\
 3\frac{f'''_1+2{\cal L}_m f'''_2}{f_2}\dot R^2 \geq 0 \,.
\end{eqnarray}

In the same way, inserting Eq. (\ref{Ricci}) into Eq. (\ref{NEC}) one obtains the NEC
\begin{equation}\label{NEC3}
\hat k \left(\hat T_\mn +T_\mn\right)k^\mu k^\nu \geq 0
\end{equation}
and hence
\begin{equation}\label{modifiedNEC}
\rho+p+(\ddot R -H\dot R)\frac{f''_1+2{\cal L}_m f''_2}{f_2}+\frac{f'''_1+2{\cal L}_m f'''_2}{f_2}\dot R^2 \geq 0 \,.
\end{equation}

It is important to realize that these results are obtained directly from the Raychaudhuri equation. However, equivalent results could have been obtained from the transformations $\rho \rightarrow \rho + \hat\rho$ and $p\rightarrow p+ \hat p$ of the SEC and NEC of general relativity.

Naturally, setting $f_2=1$ in Eqs. (\ref{modifiedSEC}) and (\ref{modifiedNEC}) one recovers the results of Ref. \cite{Santos}. Setting $f_1=R$ and $f_2=1$ one recovers SEC and NEC of general relativity (Eqs. (\ref{SECGR}) and (\ref{NECGR})).

It is interesting to note that if one imposes that gravity is repulsive, the signs of the inequalities (\ref{SEC3}) and (\ref{NEC3}) would be the same, since the signs of the conditions (\ref{SEC}), (\ref{NEC}) and (\ref{couplingIneq})
would change, and thus the above results for the SEC and the NEC would remain unaltered.

\subsection{Dominant and Weak Energy Conditions}

For the derivation of the DEC and WEC, one can extend the results obtained above and apply the transformations $\rho \rightarrow \rho + \hat \rho$ and $p\rightarrow p + \hat p$ to the DEC and WEC of general relativity since it is natural to assume that the energy conditions will not be violated when one changes from the Jordan frame to the Einstein frame (see, for instance, Ref. \cite{Atazadeh08}).

Thus, one gets for the DEC
\begin{eqnarray}\label{modifiedDEC}
\rho - p + \left(\frac{f_1}{f_2}-\frac{f'_1+2{\cal L}_m f'_2}{f_2}R\right) -\nonumber \\
(\ddot R + 5H \dot R)\frac{f''_1+2{\cal L}_m f''_2}{f_2} -
\frac{f'''_1+2{\cal L}_m f'''_2}{f_2} \dot R^2 \geq 0
\end{eqnarray}
and for the WEC
\begin{equation}\label{modifiedWEC}
\rho + \frac{1}{2}\left(\frac{f_1}{f_2}-\frac{f'_1+2{\cal L}_m f'_2}{f_2}R\right) - 3H\frac{f''_1+2{\cal L}_m f''_2}{f_2}\dot R \geq 0 \,.
\end{equation}

As one may easily check, for $f_2=1$ one obtains the results of Ref. \cite{Santos} and if one further sets $f_1=R$, Eqs. (\ref{modifiedDEC}) and (\ref{modifiedWEC}) yield $\rho-p\geq0$ and $\rho\geq0$, as expected.

\subsection{Energy Conditions for a Class of Models}

To get some insight on the meaning of the above energy conditions, one applies then to a specific type of models where $f_{1,2}$ are given by
\begin{align}
f_1(R) &=R+\epsilon R^n \nonumber \,, \\
\varphi_2(R) &=R^m \,, \label{firstModels}
\end{align}
where $\varphi_2(R)$ is defined by Eq. (\ref{f2}). Assuming a flat RW space-time, Eq.  (\ref{RWmetric}), the energy conditions can be written in the following way,
\begin{equation}\label{4EnergyConditions}
\frac{\hat \epsilon |R|^n}{1+\hat\lambda |R|^m}\left( a-\alpha_n-\frac{2\hat\lambda {\cal L}_m}{\hat \epsilon}\alpha_m|R|^{m-n} \right) \geq b
\end{equation}
where $a$, $b$ and $\alpha_{n,m}$ depend on the energy condition under study and one defines $\hat \epsilon=(-1)^n\epsilon$ and $\hat \lambda=(-1)^m\lambda$ due to the fact that for a RW metric one has $R<0$. For the SEC, one finds
\begin{equation}
a^{SEC}=-1\,, \ \ \  b^{SEC}=-(\rho+3p)\, ,
\end{equation}
\begin{eqnarray}
\alpha^{SEC}_n= -n[1+3(n-1)(\ddot R+H\dot R)R^{-2}+\nonumber \\ 3(n-1)(n-2)R^{-3}\dot R^2].
\end{eqnarray}

For the NEC, one obtains
\begin{equation}
a^{NEC}=0\,, \ \ \  b^{NEC}=-(\rho+p)\, ,
\end{equation}
\begin{equation}
\alpha^{NEC}_n= -n (n-1) \left[(\ddot R-H\dot R)R^{-2}+(n-2)R^{-3}\dot R^2 \right]\, .
\end{equation}

For the DEC, one has
\begin{equation}
a^{DEC}=1\,, \ \ \  b^{DEC}=-(\rho-p)\, ,
\end{equation}
\begin{eqnarray}
\alpha^{DEC}_n=n(1+(n-1)(\ddot R+5H\dot R)R^{-2}+\nonumber \\
(n-1)(n-2)R^{-3}\dot R^2)\, .
\end{eqnarray}

Finally, for the WEC, one gets
\begin{equation}
a^{WEC}=\frac{1}{2}\,, \ \ \  b^{WEC}=-\rho\, ,
\end{equation}
\begin{equation}
\alpha^{WEC}_n=n\left[\frac{1}{2}+3(n-1)HR^{-2}\dot R\right]
\, .
\end{equation}

Writing the derivatives of $R$ in terms of the deceleration (q), jerk (j) and snap (s) parameters,
\begin{equation}\label{RWParameters}
q=-\frac{1}{H^2}\frac{\ddot a}{a}\, , \ \ \ j=\frac{1}{H^3}\frac{\dddot a}{a}\, , \ \ \ s=\frac{1}{H^4}\frac{\ddddot a}{a}\, ,
\end{equation}
one obtains for $\alpha_n$,
\begin{widetext}
\begin{equation}
\alpha^{SEC}_n= n\left[-1+\frac{q^2+7q+j+s+4}{2(q-1)^2}(n-1)-\frac{(j-q-2)^2}{2(q-1)^3}(n-1)(n-2)\right]
\, ,
\end{equation}
\begin{equation}
\alpha^{NEC}_n= n (n-1) \left[\frac{q^2+9 q-j+s+8}{6 (q-1)^2}-\frac{ (j-q-2)^2}{6(q-1)^3}(n-2)\right]\, ,
\end{equation}
\begin{equation}
\alpha^{DEC}_n=n\left[1-\frac{q^2+3q+5j+s-4}{6(q-1)^2}(n-1)+\frac{(j-q-2)^2}{6(q-1)^3}(n-1)(n-2)\right]\, ,
\end{equation}
\begin{equation}
\alpha^{WEC}_n=n\left[\frac{1}{2}- \frac{j-q-2}{2(1-q)^2}(n-1)\right]
\,.
\end{equation}
\end{widetext}

Given these definitions, the study of all the energy conditions can be performed by satisfying the inequality (\ref{4EnergyConditions}). In Table \ref{EnergyConditionsTable} it is presented the energy conditions for models of type Eq. (\ref{firstModels}).

\begin{table*}[t]
\begin{tabular}{|c c|c|c|}
\hline
$ $& & $\hat\epsilon >0$ & $\hat\epsilon <0$\\
\hline
\multicolumn{2}{|c|}{$\hat\lambda >0$} & $\frac{2\hat\lambda{\cal L}_m}{\hat\epsilon}\leq\frac{a-\alpha_n}{\alpha_m}|R|^{n-m}$ &$\frac{2\hat\lambda{\cal L}_m}{|\hat\epsilon|}\leq\frac{\alpha_n-a}{\alpha_m}|R|^{n-m}$ \\
\hline
\multirow{2}{*}{$\hat\lambda<0$} &$1-|\hat\lambda||R|^m>0$ &$\frac{2|\hat\lambda|{\cal L}_m}{\hat\epsilon}\geq\frac{\alpha_n-a}{\alpha_m}|R|^{n-m}$  &$\frac{2|\hat\lambda|{\cal L}_m}{|\hat\epsilon|}\geq\frac{a-\alpha_n}{\alpha_m}|R|^{n-m}$ \\
\cline{2-4}
&$1-|\hat\lambda||R|^m<0$ &$\frac{2|\hat\lambda|{\cal L}_m}{\hat\epsilon}\leq\frac{\alpha_n-a}{\alpha_m}|R|^{n-m}$ & $\frac{2|\hat\lambda|{\cal L}_m}{|\hat\epsilon|}\leq\frac{a-\alpha_n}{\alpha_m}|R|^{n-m}$\\
\hline
\end{tabular}
\caption{SEC, NEC, DEC and WEC expressed by Eq. (\ref{4EnergyConditions}) with $b=0$ and where it is assumed that $\alpha_m>0$. For $\alpha_m<0$ one just needs to change the direction of the inequalities.}
\label{EnergyConditionsTable}
\end{table*}

One realizes that all the energy conditions depend on the coupling constants of the model and on the space-time under consideration. The balance between the couplings and the space-time parameters and their evolution along the history of the universe will determine whether the energy conditions are satisfied or not. Setting the estimated values of $q$, $j$ and $s$ from observation, one can impose bounds on the coupling constants of the gravity model (see e.g. Refs. \cite{Santos, Atazadeh08, PerezBergliaffa:2006ni} for discussions).

When applying for models like Eq. (\ref{firstModels}), the condition for attractive gravity (AG), Eq. (\ref{couplingIneq}), gives, once again, an inequality like Eq. (\ref{4EnergyConditions}) with
\begin{equation}
a^{AG}=1\, ;\ \ \  b^{AG}=0\, ,
\end{equation}
\begin{equation}
\alpha^{AG}_n= -\frac{n}{6H^2(1-q)}\, .
\end{equation}
Thus, the results depicted  in Table \ref{EnergyConditionsTable} also stand for the condition that ensures that gravity remains attractive.

\section{Generalized Dolgov-Kawasaki Instability}\label{massInstability}

Modified gravity theories may be subjected to instabilities. These are due to the fact that the field equations (\ref{fieldEq}) are of fourth order and their trace give a dynamical equation for $R$. Such instabilities do not occur in general relativity since there the field equations are of second order and their trace give an algebraic equation for $R$. In order for the dynamical field $R$ to be stable its ``mass'' must be positive, a requirement usually referred to as Dolgov-Kawasaki stability criterion.

The Dolgov-Kawasaki criterion was studied in the context of $f(R)$ theories in Ref. \cite{Nojiri03} and generalized for the models under study here, i.e. models whose action is given by Eq. (\ref{action}),  in Ref. \cite{Faraoni} and can be expressed as
\begin{equation}\label{inequalityDK}
f_1''(R)+2\lambda{\cal L}_m\varphi''_2(R)\geq0
\end{equation}
where $\varphi_2(R)$ is defined by Eq. (\ref{f2}) and $f_1(R)$ is written as $f_1(R)=R+\epsilon\varphi_1(R)$, with $\epsilon$ being a constant. Notice that  Eq. (\ref{inequalityDK}) is not quite the one presented in Ref. \cite{Faraoni} given that there the Lagrangian density, ${\cal L}_m$, was not considered.

For models such as the ones defined by Eq. (\ref{firstModels}), the inequality (\ref{inequalityDK}) becomes
\begin{equation}\label{inequality1}
\hat\epsilon n(n-1) |R|^{n-2} +2\hat\lambda{\cal L}_m m(m-1)|R|^m\geq0 \,,
\end{equation}
where $\hat\epsilon$ and $\hat\lambda$ are defined by,
\begin{equation}
\hat\epsilon={\left\{\begin{array}{cc} (-1)^n\epsilon,&\mbox{ if } R< 0\\
\epsilon, & \mbox{ if } R>0\end{array}\right.}\,, \ \ \
\hat\lambda={\left\{\begin{array}{cc} (-1)^m\lambda,&\mbox{ if } R< 0\\
\lambda, & \mbox{ if } R>0\end{array}\right.}\,.
\end{equation}

Notice that this inequality can also be written as Eq. (\ref{4EnergyConditions}) for,
\begin{equation}
a^{DK}=0\,,\ \ \  b^{DK}=0\, ,
\end{equation}
\begin{equation}
\alpha^{DK}_n=-\frac{n(n-1)}{36H^4(1-q)^2}(1+\hat\lambda|R|^m)\, .
\end{equation}

Thus, the results presented in Table \ref{EnergyConditionsTable}, for the cases with $1+\hat\lambda|R|^m>0$, stand for Eq. (\ref{inequality1}) too. However, it is important to realize that although the energy conditions and the Dolgov-Kawasaki criterion yield the same type of inequalities, they are independent from each other. From Table \ref{EnergyConditionsTable}, one can see that, if the factors $\frac{a-\alpha_n}{\alpha_m}$  are the same, the various conditions (the four energy conditions, the condition for attractive gravity and the Dolgov-Kawasaki criterion) are degenerate.

For most of the cases, the viability of the model with respect to the Dolgov-Kawasaki instability criterion will depend not only on the value of the constants $\epsilon$ and $\lambda$ (which may be further constrained from the solar system observations \cite{Bertolami:2008im}), but also on the space-time metric under consideration.

>From the Dolgov-Kawasaki criterion one has
\begin{equation}
\frac{\alpha^{DK}_n}{\alpha^{DK}_m}=\frac{n(n-1)}{m(m-1)}\,,
\end{equation}
and thus the stability criterion for models defined by Eq. (\ref{firstModels}) do not depend on the values of $j$ or $s$, and the space-time dependence is present only through $R$. Furthermore, since $\alpha^{DK}_n/\alpha^{DK}_m>0$, one can see that models for which $\hat\lambda>0$ and $\hat\epsilon>0$, the Dolgov-Kawasaki criterion is always satisfied independently of the values of $\epsilon$, $\lambda$ and $R$ (note that as $\alpha^{DK}_m<0$, the direction of the inequalities in Table \ref{EnergyConditionsTable} are reversed).

For models where $m=n$ the inequality Eq. (\ref{inequality1}) implies, $\epsilon + 2\lambda{\cal L}_m \geq 0$.

Another type of models that is interesting to consider are the ones where $f_{1,2}(R)$ are given by
\begin{align}
f_1(R) &=\sum^k_{n=1}a_n R^n \nonumber \\
\varphi_2(R) &=\sum^{k'}_{m=1}b_mR^m
\end{align}

For these models, the inequality Eq. (\ref{inequalityDK}) becomes,
\begin{equation}\label{inequality2}
\sum^k_{n=2}a_n n(n-1)R^{n-2}+2\lambda{\cal L}_m \sum^{k'}_{m=2}b_m m(m-1)R^{m-2}\geq0
\end{equation}

The stability conditions for models with $k,k'=2,3$ are presented in Table \ref{secondModelTable}.
%
\begin{table*}[t]
\begin{tabular}{|c|c|c|c|c|}
\hline
& \multicolumn{2}{|c|}{$k'=2$} & \multicolumn{2}{|c|}{$k'=3$}\\
\hline
\multirow{2}{*}{$k=2$} &\multicolumn{2}{c|}{\multirow{2}{*}{ $a_2+2\lambda{\cal L}_m b_2\geq 0$}}&$a_3>0$&$R\geq-\frac{a_2+2\lambda{\cal L}_m b_2}{3 a_3}$\\
\cline{4-5}
&\multicolumn{2}{c|}{} &$a_3<0$&$R\leq \frac{a_2+2\lambda{\cal L}_m b_2}{3 |a_3|}$\\
\hline
\multirow{2}{*}{$k=3$} &$\lambda{\cal L}_m b_3>0$&$R\geq-\frac{a_2+2\lambda{\cal L}_m b_2}{6 \lambda{\cal L}_m b_3}$&$a_3+2\lambda{\cal L}_m b_3>0$&$R\geq-\frac{a_2+2\lambda{\cal L}_m b_2}{3(a_3+2 \lambda{\cal L}_m b_3)}$\\
\cline{2-5}
&$\lambda{\cal L}_m b_3<0$&$R\leq \frac{a_2+2\lambda{\cal L}_m b_2}{6 |\lambda{\cal L}_m b_3|}$&$a_3+2\lambda{\cal L}_m b_3<0$&$R\leq \frac{a_2+2\lambda{\cal L}_m b_2}{3|a_3+2 \lambda{\cal L}_m b_3|}$\\
\hline
\end{tabular}
\caption{Stability conditions for models defined by Eq. (\ref{inequality2}), where it is assumed that $R>0$. }
\label{secondModelTable}
\end{table*}
%

Once again, one sees that the stability conditions for most of the models are metric dependent. This fact is in strong contrast with what happens for $f(R)$ theories where the stability can be ensured independently of the space-time. This difference can  be relevant and it brings about further constraints on $f(R)$ theories with a matter-curvature coupling since it restricts the Ricci scalar,  which might turn out to be unphysical.

\section{Conclusions and Outlook}\label{conclusions}

In this work, the well known energy conditions and the Dolgov-Kawasaki stability criterion, for the modified theories of gravity described by action (\ref{action}) were examined. In section \ref{energyConditions} it was shown that the generalized SEC and NEC can be obtained directly from the SEC and NEC of general relativity through the transformations $\rho\rightarrow\rho+\hat \rho$ and $p\rightarrow p + \hat p$, where the quantities $\hat \rho$ and $\hat p$ contain the physics of the higher-order terms in $R$. The same procedure can be applied to the DEC and the WEC. As expected, all generalized energy conditions for a suitable energy-momentum tensor are strongly dependent on the geometry. Conditions to keep the effective gravitational coupling positive and gravity attractive (Eq. (\ref{couplingIneq})) were also obtained. The encountered energy conditions have been applied to a specific kind of models and shown that all the energy conditions together with the positiveness o
 f  the effective gravitational constant and the Dolgov-Kawasaki criterion can be expressed by the same type of inequalities.

In section \ref{massInstability} the Dolgov-Kawasaki stability criterion was used to test some modified gravity models. One finds that the stability of a specific model will depend not only on its couplings but also on the space-time geometry in question as depicted in Table \ref{secondModelTable}.

We stress that our method and in particular Eqs. (\ref{couplingIneq}), (\ref{modifiedSEC}), (\ref{modifiedNEC}), (\ref{modifiedDEC}) and (\ref{modifiedWEC}) are quite general and can be used to study the physical implications of any $f(R)$ model with non-minimal curvature-matter coupling




\begin{thebibliography}{99}

\bibitem{Nojiri06} S. Nojiri and S. D.
Odintsov, hep-th 0601213.

\bibitem{Sotiriou:2008rp}
  T.~P.~Sotiriou and V.~Faraoni,
  arXiv:0805.1726 [gr-qc].

\bibitem{Bertolami:2007}
  O.~Bertolami, C.~G.~Boehmer, T.~Harko and F.~S.~N.~Lobo,
  Phys.\ Rev.\  D {\bf 75} (2007) 104016
  [arXiv:0704.1733 [gr-qc]].

\bibitem{Bertolami:2008zh}
  O.~Bertolami, T.~Harko, F.~S.~N.~Lobo and J.~P\'aramos,
  arXiv:0811.2876 [gr-qc].

\bibitem{Odintsov}
  S.~Nojiri and S.~D.~Odintsov,
  Phys.\ Lett.\  B {\bf 599}, 137 (2004)
  [arXiv:astro-ph/0403622];
  G.~Allemandi, A.~Borowiec, M.~Francaviglia and S.~D.~Odintsov,
  Phys.\ Rev.\  D {\bf 72}, 063505 (2005)
  [arXiv:gr-qc/0504057];
  S.~Nojiri and S.~D.~Odintsov,
  eConf {\bf C0602061}, 06 (2006)
  [Int.\ J.\ Geom.\ Meth.\ Mod.\ Phys.\  {\bf 4}, 115 (2007)]
  [arXiv:hep-th/0601213].

\bibitem{Sotiriou08} T.~P.~Sotiriou, Phys.\ Lett.\  B {\bf 664}, 225 (2008) 
[arXiv:0805.1160v2 [gr-qc]].

\bibitem{Sotiriou:2008it}
  T.~P.~Sotiriou and V.~Faraoni,
  Class.\ Quant.\ Grav.\  {\bf 25} (2008) 205002
  [arXiv:0805.1249 [gr-qc]].

\bibitem{Bertolami:2008ab}
  O.~Bertolami, F.~S.~N.~Lobo and J.~P\'aramos,
  Phys.\ Rev.\  D {\bf 78} (2008) 064036
  [arXiv:0806.4434 [gr-qc]].

\bibitem{Puetzfeld:2008xu}
  D.~Puetzfeld and Y.~N.~Obukhov,
  Phys.\ Rev.\  D {\bf 78} (2008) 121501
  [arXiv:0811.0913 [astro-ph]].

\bibitem{Bertolami:2008im}
  O.~Bertolami and J.~P\'aramos,
  Class.\ Quant.\ Grav.\  {\bf 25} (2008) 245017
  [arXiv:0805.1241 [gr-qc]].

\bibitem{Bertolami:2007vu}
  O.~Bertolami and J.~P\'aramos,
  Phys.\ Rev.\  D {\bf 77}, 084018 (2008)
  [arXiv:0709.3988 [astro-ph]].

\bibitem{Faraoni}
  V.~Faraoni,
  Phys.\ Rev.\  D {\bf 76} (2007) 127501
  [arXiv:0710.1291 [gr-qc]].

\bibitem{Hawking:1973uf}
  S.~W.~Hawking and G.~F.~R.~Ellis,
  ``The Large scale structure of space-time,''
{\it  (Cambridge University Press, Cambridge  1973)}.

\bibitem{Santos}
	J.~Santos, J.~S.~Alcaniz, M.~J.~Rebou\c cas and F.~C.~Carvalho,
  Phys.\ Rev.\  D {\bf 76} (2007) 083513
  [arXiv:0708.0411 [astro-ph]].

\bibitem{Atazadeh08}  K. Atazadeh, A. Khaleghi, H. R. Sepangi, Y. Tavakoli,
arXiv:0811.4269 [gr-qc].

\bibitem{PerezBergliaffa:2006ni}
  S.~E.~Perez Bergliaffa,
  Phys.\ Lett.\  B {\bf 642}, 311 (2006)
  [arXiv:gr-qc/0608072].

\bibitem{Nojiri03}
  S.~Nojiri and S.~D.~Odintsov,
  Phys.\ Rev.\  D {\bf 68}, 123512 (2003)
  [arXiv:hep-th/0307288].








\end{thebibliography}
\end{document}